\documentclass[prb,longbibliography,showpacs,twocolumn,superscriptaddress,amsmath,amssymb,verbatim]{revtex4-2}

\usepackage{graphicx}% Include figure files
\usepackage{lmodern}
\usepackage{epstopdf}
\usepackage{dcolumn}% Align table columns on decimal point
\usepackage{bm}% bold math
\usepackage{subfigure}
\usepackage{amsmath} %math

\usepackage[pdftex,colorlinks=true,citecolor=blue,linkcolor=blue,urlcolor=blue,bookmarks=true]{hyperref}
\usepackage[sort&compress]{natbib}

\usepackage{color}
\usepackage{textcomp}
\definecolor{red}{rgb}{1,0,0}

\definecolor{blue}{rgb}{0,0,1}

\definecolor{green}{rgb}{0,1,0}

\begin{document}
\preprint{APS}

\title{Competing magnetic phases in the frustrated spin-1/2 chain compound $\beta$-TeVO$_4$ probed by NMR}

\author{M. Pregelj}
\email{matej.pregelj@ijs.si}
\affiliation{Jo\v{z}ef Stefan Institute, Jamova cesta 39, 1000 Ljubljana, Slovenia}
\author{A. Zorko}
\affiliation{Jo\v{z}ef Stefan Institute, Jamova cesta 39, 1000 Ljubljana, Slovenia}
\affiliation{Faculty of Mathematics and Physics, University of Ljubljana, Jadranska u. 19, 1000 Ljubljana, Slovenia}
\author{D. Ar\v{c}on}
\affiliation{Jo\v{z}ef Stefan Institute, Jamova cesta 39, 1000 Ljubljana, Slovenia}
\affiliation{Faculty of Mathematics and Physics, University of Ljubljana, Jadranska u. 19, 1000 Ljubljana, Slovenia}
\author{M. Klanj\v{s}ek}
\affiliation{Jo\v{z}ef Stefan Institute, Jamova cesta 39, 1000 Ljubljana, Slovenia}
\author{N. Jan\v{s}a}
\affiliation{Jo\v{z}ef Stefan Institute, Jamova cesta 39, 1000 Ljubljana, Slovenia}
\author{P. Jegli\v{c}}
\affiliation{Jo\v{z}ef Stefan Institute, Jamova cesta 39, 1000 Ljubljana, Slovenia}
\author{O. Zaharko}
\affiliation{Laboratory for Neutron Scattering and Imaging, PSI, CH-5232 Villigen, Switzerland}
\author{S. Kr\"{a}mer}
\affiliation{Laboratoire National des Champs Magn\'etiques Intenses, LNCMI-CNRS (UPR3228), EMFL, Universit\'e \\ Grenoble Alpes, UPS and INSA Toulouse, Bo\^{i}te Postale 166, 38042 Grenoble Cedex 9, France}
\author{M. Horvati\'{c}}
\affiliation{Laboratoire National des Champs Magn\'etiques Intenses, LNCMI-CNRS (UPR3228), EMFL, Universit\'e \\ Grenoble Alpes, UPS and INSA Toulouse, Bo\^{i}te Postale 166, 38042 Grenoble Cedex 9, France}
\author{A. Prokofiev}
\affiliation{Institute of Solid State Physics, Vienna University of Technology, Wiedner Hauptstrasse 8-10, 1040 Vienna, Austria}

\date{\today}

\begin{abstract}

In frustrated spin-1/2 chains the competition between the nearest- and next-nearest-neighbor exchange interactions leads to a rich phase diagram that becomes even richer in the presence of perturbations in their material realizations.
These effects are still largely unexplored, so that new insight into static and dynamic magnetism, in particular by sensitive local probes, is highly desired.
Here we present a comprehensive $^{17}$O nuclear magnetic resonance study of $\beta$-TeVO$_4$, where the anisotropy of the main exchange interactions and additional weak interchain exchange interactions complement the theoretical phase diagram.
Our results confirm the dynamical nature of the intriguing spin-stripe phase that has been reported in previous studies.
In addition, we find that the magnetic order in the high-field phase, which develops just below the magnetization saturation, is consistent with an unusual type of spin-density-wave (SDW) order with different alignments of the magnetic moments on the neighboring chains.
This is reminiscent of the ordering in the SDW phase, realized in the absence of the magnetic field, and is thus most likely stabilized by magnetic anisotropy.

\end{abstract}

\maketitle

\section{Introduction}

Frustrated spin-1/2 chains, i.e., chains with nearest- and next-nearest-neighbor Heisenberg exchange interactions, $J_1$ and $J_2$, respectively, have been extensively studied \cite{castilla1995quantum, seki2008correlation, nishimoto2011saturation, cemal2018field} due to their rich phase diagram \cite{hikihara2008vector, sudan2009emergent, hikihara2010magnetic}.
Recently, the focus has been on compounds with ferromagnetic $J_1$ \cite{mourigal2012evidence, willenberg2016complex, orlova2017nuclear, cemal2018field}, where a spin-nematic state, exhibiting long-range order of magnetic quadrupoles born out of magnon pairing \cite{chubukov1991chiral, shannon2006nematic, zhitomirsky2010magnon}, has been predicted \cite{hikihara2008vector, sudan2009emergent, hikihara2010magnetic}. 
Moreover, besides this enigmatic state, ferromagnetic frustrated spin-1/2 chains show a plethora of other intriguing phases \cite{hikihara2008vector, sudan2009emergent}.
To begin with, the zero-field magnetic ground state is characterized by long-range order of vector-chiral (VC) correlations.
With increasing the magnetic field, a spin-density-wave (SDW) phase develops, which is characterized by incommensurate amplitude modulation of the ordered magnetic moments.
Just below the magnetization saturation, the emergence of the spin-nematic state is anticipated,
but its realization is in real systems typically hindered by perturbations, e.g., small additional intra- or inter-chain interactions or magnetic anisotropy \cite{cemal2018field, bush2018exotic, pregelj2019magnetic}.
On the other hand, perturbations may lead to novel phases with intricate magnetic properties that further enrich the phase diagram of the frustrated spin-1/2 chain \cite{ueda2020roles} and are yet to be explored experimentally.

\begin{figure}[!]
\centering
\includegraphics[width=\columnwidth]{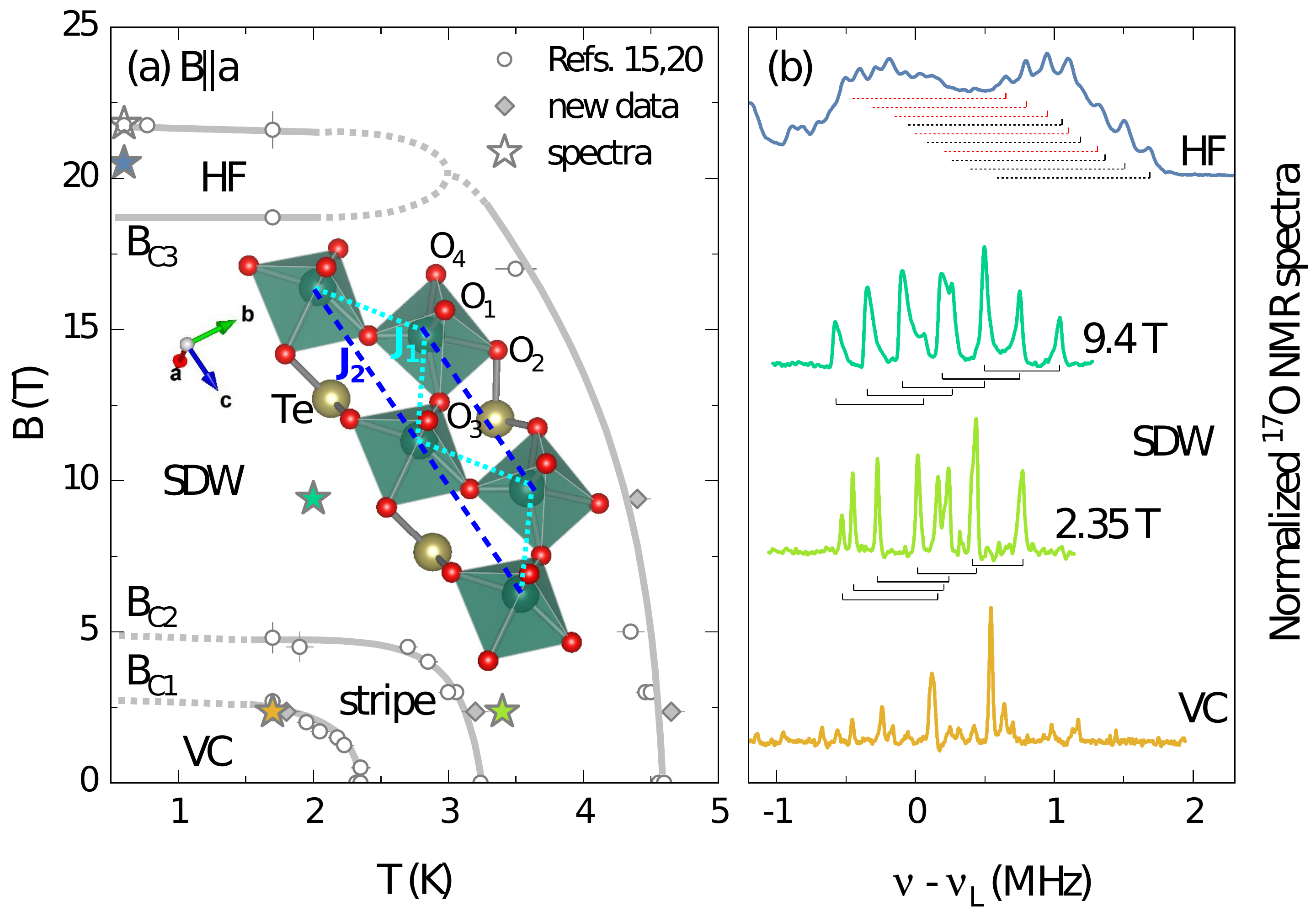}
\caption{(a) Magnetic phase diagram of $\beta$-TeVO$_4$ in magnetic field applied along the $a$ crystallographic axis. Inset: a part of the V$^{4+}$ ($S$\,=\,1/2) zigzag chain, consisting of corner sharing VO$_5$ pyramids that include four crystalographically different O sites. $J_1$ is bridged by O atoms, whereas $J_2$ is bridged via O-Te-O paths. (b) $^{17}$O NMR spectra for the highest-frequency O site measured at different positions in the phase diagram [marked by corresponding stars of the same color in (a)]. The horizontal square brackets denote the U singularities corresponding to the same satellite line split by the incommensurate SDW order. Only one side of the bracket can be resolved in the HF phase. }
\label{fig-PD}
\end{figure}

An intriguing compound containing zigzag chains of magnetic V$^{4+}$ ($S$\,=\,1/2) ions is $\beta$-TeVO$_4$ \cite{meunier1973oxyde}.
This system exhibits a magnetic phase diagram [Fig.\,\ref{fig-PD}(a)] that matches very well the theoretical phase diagram of $J_1$-$J_2$ spin-1/2 chain with ferromagnetic $J_1$\,$\sim$\,$-38$\,K and antiferromagnetic $J_2$\,$\sim$\,$-J_1$ \cite{pregelj2015spin}.
Yet, the exchange anisotropy and weak interchain interactions lead to some unexpected properties.
The system orders below $T_{N1}$\,=\,4.6\,K, where the SDW phase develops, whereas the VC magnetic ground state is established below $T_{N3}$\,=\,2.3\,K. 
Between $T_{N2}$\,=\,3.3\,K and $T_{N3}$, however, an extraordinary spin-stripe phase emerges, where the two SDW order parameters that constitute the VC phase have different magnetic ordering vectors \cite{pregelj2015spin}.
This phase is characterized by unusual spin dynamics in the MHz frequency range detected by muon spin relaxation \cite{pregelj2019elementary}, which still lacks an independent experimental confirmation.
Applying the magnetic field at lowest temperatures induces the same spin-stripe phase between $B_{C1}$\,$\approx$\,2.5\,T and $B_{C2}$\,$\approx$\,5\,T, which is followed by the SDW phase at higher fields (Fig.\,\ref{fig-PD}).
In the SDW phase the magnetic ordering vector monotonically decreases with increasing magnetic field \cite{pregelj2019magnetic}, which complies with theoretical predictions \cite{sudan2009emergent}. 
However, this dependence is broken at $B_{C3}$\,$\approx$\,18\,T, where, instead of the anticipated spin-nematic phase, a high-field (HF) phase with yet unknown incommensurate dipolar magnetic order has been found \cite{pregelj2019magnetic,pregelj2020thermal}.

Here we explore the magnetic phase diagram of $\beta$-TeVO$_4$ by $^{17}$O nuclear magnetic resonance (NMR). 
Previous low-magnetic-field NMR study \cite{weickert2016magnetic} that has been focused on $^{125}$Te nuclei was not able to probe the magnetism below $\sim$10\,K.
This was due to direct involvement of the Te ions in the main superexchange paths, leading to a very strong hyperfine interaction and consequent extremely rapid relaxation.
In contrast to Te, one of four crystallographically inequivalent O sites experiences sufficiently small hyperfine fields, allowing us to successfully probe the NMR response also in the long-range-ordered magnetic phases.
In fact, by extending the measurement range down to 0.5\,K and up to 21.8\,T, we are able to confirm the dynamical nature of the spin-stripe phase, reveal details about the high-field magnetic structure, and observe potential magnetic-field-induced closing of the magnon gap in the SDW phase.
Our study, hereby, represents an important complementary insight into the magnetic properties of $\beta$-TeVO$_4$ and illuminates the influence of anisotropic interactions on magnetic phases in the $J_1$-$J_2$ spin-1/2 chain compounds from a local perspective.

\section{Experimental}

The same single-crystal sample (1.00$\times$3.10$\times$7.55\,mm$^3$) was used as in our previous NMR study \cite{pregelj2020thermal}. 
The crystal was grown by chemical vapor transport reaction using TeO$_2$ and VO$_2$ as starting materials and TeCl$_4$ as a transport agent.
The enrichment of 8\% $^{17}$O in the final product was achieved by a preceding partial oxidation of V$_2$O$_3$ to VO$_2$ by $^{17}$O$_2$.
The $^{17}$O NMR was measured using a custom-built spectrometer at Jo\v{z}ef Stefan Institute, Slovenia, as well as a high-field spectrometer at the Laboratoire National des Champs Magnétiques Intenses (LNCMI), Grenoble, France, altogether covering the magnetic field, $B$, range between 2.3 and 24\,T.
The shape of the coil was adapted to fit the crystal (flat elongated plate), while a goniometer was used to adjust the orientation of the sample with respect to the magnetic field direction, which was limited to the crystallographic $ab$ plane.
The transverse magnetization, $M_\perp$, decay controlled by the spin-spin relaxation $1/T_2$ was measured with a standard Hahn-echo pulse sequence and fitted by a single-exponential function $\exp(-2\tau/T_2)$, where $\tau$ is the time between the $\pi/2$ and $\pi$ pulses.
The decay of the longitudinal magnetization, $M_z$, driven by the spin-lattice relaxation $1/T_1$ was fitted to the expression for the outer-satellite line of a spin $I$\,=\,5/2 nucleus \cite{simmons1962nuclear}:
\begin{equation}
\begin{split}
M_z(t) = &\, M_0 - M_1 [1/35\exp(-t/T_1)\\
&+3/14\exp(-3t/T_1)+2/5\exp(-6t/T_1)\\
&+2/7\exp(-10t/T_1)+1/14\exp(-15t/T_1) ],
\label{T1function}
\end{split}
\end{equation}
where $t$ is the time after the initial inversion pulse, $M_0$\,=\,$M_z(t$\,$\to$\,$\infty)$ the equilibrium magnetization and $M_1$\,=\,$M_0 - M_z(t$\,=\,0) the amplitude of the magnetization inversion.

\section{Results}

\begin{figure}[!]
\centering
\includegraphics[width=\columnwidth]{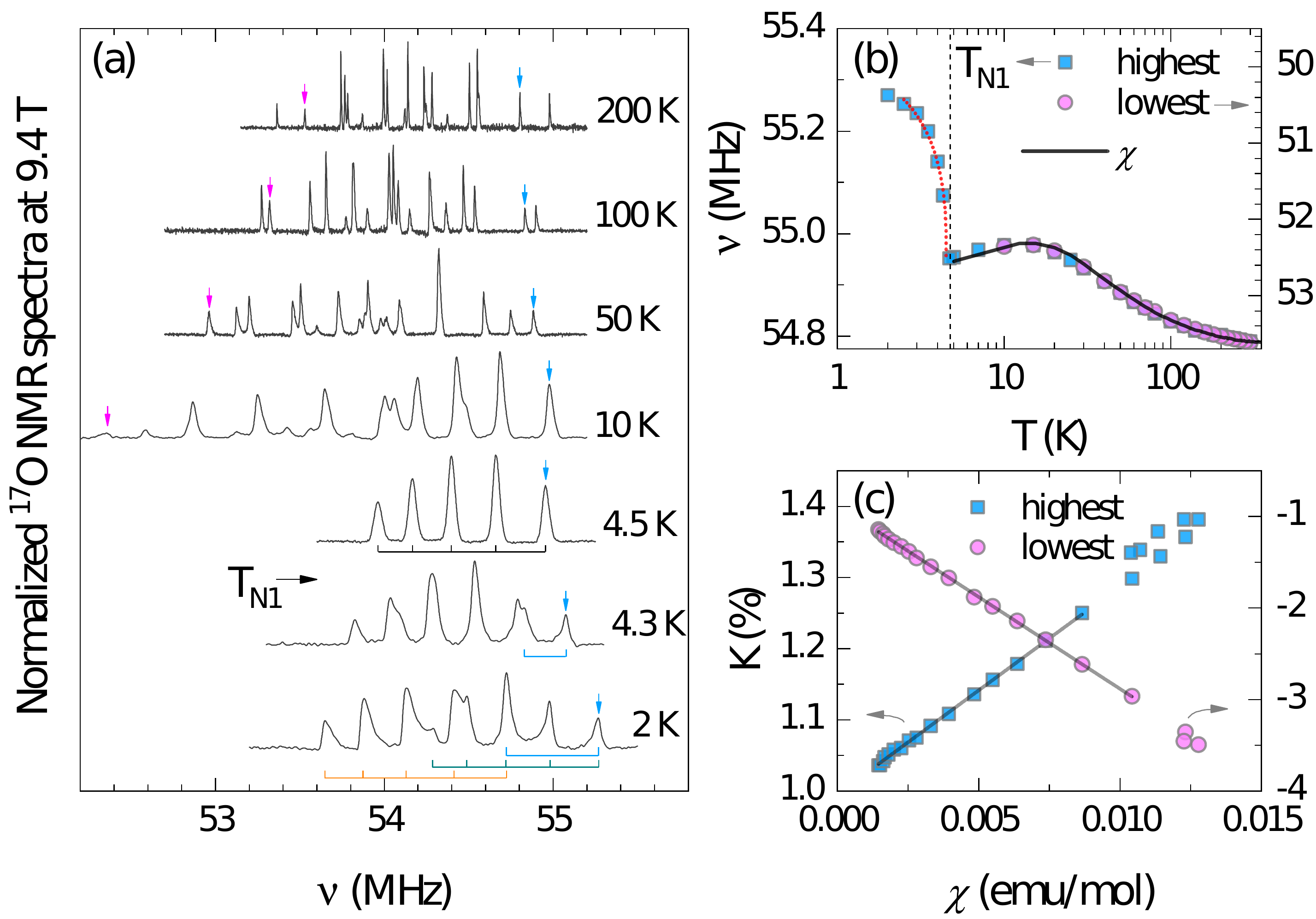}
\caption{(a) The temperature dependence of the $^{17}$O NMR spectrum measured at 9.4\,T. The horizontal square brackets denote the U singularities corresponding to the highest-frequency satellite. Five-pointed forks denote individual quintets obtained by fit (see text for details). (b) Shift of the lowest- and highest-frequency satellite lines (symbols), marked by arrows in (a), in comparison to magnetic susceptibility data from Ref.\,[\onlinecite{pregelj2015spin}]. The dotted line represents the  $(T-T_{N1})^\beta$ model with $\beta$\,=\,0.37(3). (c)  Clogston-Jaccarino plot for the lowest- and highest-frequency satellite lines above $T_{N1}$. Solid lines represent linear fits.}
\label{fig-400}
\end{figure}
To determine the coupling between the $^{17}$O nuclei and the magnetic moments of the magnetic ions as well as to obtain an overview of the magnetic behavior in the paramagnetic and SDW states, we first measured temperature dependence of the $^{17}$O NMR spectrum of $\beta$-TeVO$_4$ in the magnetic field $B$\,=\,9.4\,T applied along the crystallographic $a$ axis, i.e., perpendicular to the VO$_5$ chains.	
At room temperature, the spectrum comprises twenty sharp peaks [Fig.\,\ref{fig-400}(a)] corresponding to $^{17}$O ($I$\,=\,5/2) quintets from four crystallographically inequivalent O sites split by quadrupole interaction.
We note that some of the spectral lines overlap. 
On cooling three quintets shift to lower frequencies while the fourth quintet shifts to higher frequencies, i.e., in agreement with our high-filed study \cite{pregelj2020thermal}.
This response becomes obvious when plotting the temperature dependence of the positions of two representative spectral lines [Fig.\,\ref{fig-400}(b)] marked by arrows in Fig.\,\ref{fig-400}(a).
The two shifts are significantly different, as the line moving to lower frequencies shifts $\sim$7-times more than the one shifting to higher frequencies, indicating substantially weaker hyperfine coupling for the latter. 
Both shifts exhibit a pronounced maximum at $\sim$12.7\,K, perfectly matching the temperature dependence of the magnetic susceptibility $\chi$ [Fig.\,\ref{fig-400}(b)].
In fact, plotting the magnetic hyperfine, i.e., Knight, shift $K$\,=\,$(\nu-\nu_L)/\nu_L$, where $\nu_L$\,=\,$^{17}\gamma B$ is the Larmor frequency, $^{17}\gamma$\,=\,5.772\,MHz/T  is the $^{17}$O gyromagnetic ratio and $B$ is the applied magnetic field, as a function of $\chi$, i.e., the so called Clogston-Jaccarino plot  [Fig.\,\ref{fig-400}(c)], we find a clear linear dependence, where the slope is directly proportional to the strength of the hyperfine interactions \cite{clogston1964interpretation}.
The derived hyperfine couplings for $B||a$ are $A_{\text{low}}$\,=\,$-$1.12(2)\,T/$\mu_B$ and $A_{\text{high}}$\,=\,0.163(2)\,T/$\mu_B$ for the low and the high frequency site, respectively.
The bigger value is comparable with the values derived for the $^{125}$Te NMR amounting up to 5.8\,T/$\mu_B$ \cite{weickert2016magnetic}, leading to a similar loss of the NMR signal below 10 K.
The other coupling value is $\sim$7 times smaller, which explains why this site could be used for the low-temperature NMR study.

\begin{figure}[!]
\centering
\includegraphics[width=\columnwidth]{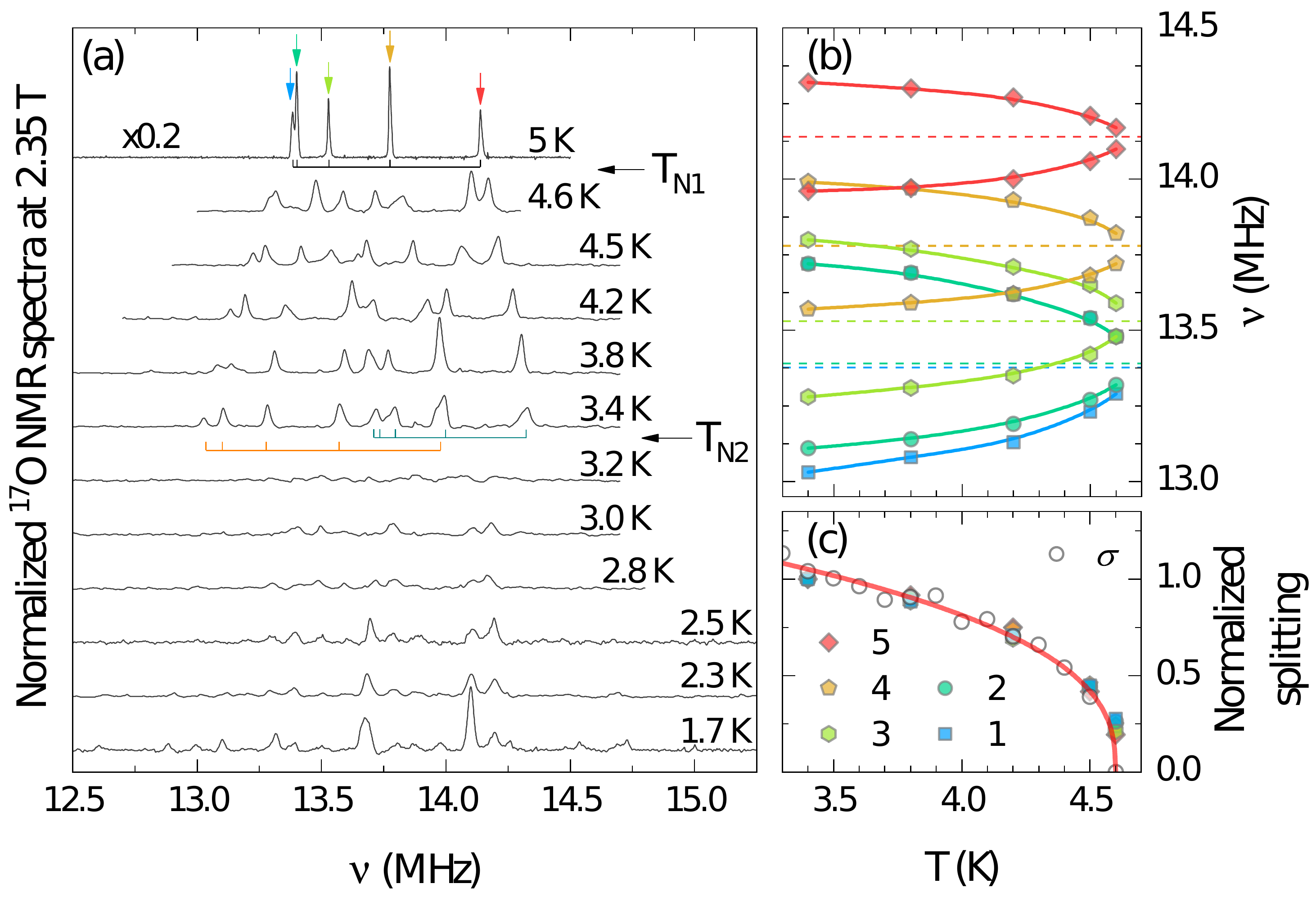}
\caption{(a) The temperature dependence of the $^{17}$O NMR spectrum measured at 2.35\,T at low temperatures. Five-pointed forks denote individual quintets obtained by fit (see text for details).  (b) The peak frequencies demonstrating the splitting of the five spectral lines due to the establishment of the incommensurate magnetic order. Solid lines are guides to the eye. Colored dashed lines mark the positions of the spectral lines in the paramagnetic state at 5\,K marked by arrows in (a). (c) The temperature dependence of the normalized NMR splittings (solid symbols) compared to the magnetic order parameter $\sigma$ derived from neutron diffraction (empty symbols) \cite{pregelj2019elementary}. The solid line is the fit of the $(T-T_{N1})^\beta$ model, with $\beta$\,=\,0.37(3). }
\label{fig-100}
\end{figure}

On further cooling, the spectral lines shifting to lower frequencies loose intensity and finally disappear below $\sim$7\,K [Fig.\,\ref{fig-400}(a)], which is due to extremely fast spin-spin ($T_2$) relaxation driven by stronger hyperfine couplings of these lines compared to the lines that shift to higher frequencies.
Consequently, at 4.5\,K, i.e., just above the magnetic transition, the NMR spectrum reduces to only five lines, namely, a single quintet corresponding to the O site with the weakest hyperfine coupling.
This particular O site is most likely the apical O$_1$ site that is isolated from the main exchange pathways [inset in Fig.\,\ref{fig-PD}(a)].
Below $T_{N1}$, each of the five lines broadens into a U-shaped spectrum [Fig.\,\ref{fig-400}(a)], characteristic of incommensurate amplitude-modulated magnetic order in the SDW phase with the width of the U splitting determined by the amplitude of the projection of the magnetic-moments modulation on the external magnetic field.
The position of the outer satellite line below $T_{N1}$ [Fig.\,\ref{fig-400}(b)], therefore, reflects the amplitude of the ordered magnetic moments in the SDW phase, which scales with temperature as $(T-T_{N1})^\beta$, yielding $\beta$\,=\,0.37(3) [see Figs.\,\ref{fig-400}(b) and \ref{fig-100}(c)].
The latter value is in agreement with the critical exponents for the order parameter in three-dimensional (3D) spin lattices, ranging from 0.33 to 0.36 for the Ising and Heisenberg case, respectively, \cite{pelissetto2002critical, chaikin2000principles}.
Indeed, the 3D nature of critical correlations is in line with the existence of sizable interchain, i.e., 3D, interactions identified in previous studies \cite{saul2014density, pregelj2018coexisting}.

To explore all the low-field ordered phases in the same applied field, we performed additional $^{17}$O NMR measurements at 2.35\,T.
The temperature dependence of the spectrum measured below 5\,K is shown in Fig.\,\ref{fig-100}(a).
The relative positions of the five spectral lines of the remaining quintet are significantly different from those collected at 9.4\,T.
This is a consequence of the $^{17}$O Larmor frequency at 2.35\,T being strongly affected by the quadrupolar interaction \cite{bain2004from}, determined by the $^{17}$O quadrupolar moment and the local electric-field-gradient (EFG) tensor.
Nevertheless, below $T_{N1}$, similarly as for measurements at 9.4\,T, also at 2.35\,T each of the five lines splits into two that monotonically shift away from each other on cooling [Fig.\,\ref{fig-100}(b)], as expected for the SDW, i.e., amplitude-modulated, phase \cite{pregelj2013evolution}.
In fact, the temperature dependence of the U-splitting [Fig.\,\ref{fig-100}(c)] follows the same $(T-T_{N1})^\beta$ dependence as the outer satellite at 9.4\,T [Fig.\,\ref{fig-400}(b)], while both dependencies exactly match the evolution of the magnetic order parameter obtained from neutron diffraction at zero field [Fig.\,\ref{fig-100}(c)] \cite{pregelj2019elementary}.

A detailed inspection of the U-split lines in the SDW phase [Fig.\,\ref{fig-PD}(b)] provides further insight into quadrupolar effects.
Namely, while the U splittings of different lines in the quintets are rather similar at 9.4\,T, they differ significantly at 2.35\,T, yet again indicating that quadrupolar effects at 2.35\,T are strong and that first-order perturbation theory no longer applies.
It is thus convenient to consider the edge singularities of the U-shaped lines as two quintets corresponding to two magnetically-inequivalent O$_1$ sites; one that on cooling shifts to higher and one that shifts to lower frequencies [Fig.\,\ref{fig-400}(a) and \ref{fig-100}(a)].
The observed behavior can then be described by exact diagonalization of the quadrupolar Hamiltonian $H=\hbar \gamma {\bf I}\cdot {\bf B_{\text{eff}}}+h \nu_Q [3I_z^2-I(I+1)+\eta(I_+^2+I_-^2 )/2]/6$, where $h$ is the Planck constant, $\hbar$\,=\,$h/2\pi$, {\bf I}\,=\,$(I_x,\,I_y,\,I_z)$ is the nuclear spin vector, $I_\pm$\,=\,$I_x$\,$\pm$\,$iI_y$, {\bf B}$_{\text{eff}}$ is the effective magnetic field, $\nu_Q$ is the quadrupolar frequency, and $\eta$ is the EFG asymmetry parameter \cite{aimo2009spin}.
Adjusting $\nu_Q$, $\eta$ as well as the orientation and the size of {\bf B}$_{\text{eff}}$, we can exactly reproduce the two quintets measured in the paramagnetic phase, i.e., at 5\,K, in the field of 2.35 and 9.4\,T as well as the four (two in 2.35\,T and two in 9.4\,T) 
quintets in the SDW phase [five-pointed forks in Fig.\,\ref{fig-400}(a) and \ref{fig-100}(a)].
The derived parameters are $\nu_Q$\,=\,1.05(5)\,MHz, $\eta$\,=\,0.62(5), and the orientations of the two principal EFG axes pointing along (0.7,~0,~$-$0.7) and (0.5,~0.7,~0.5) in the $a^*bc$ crystalographic system, which is in line with the fact that the O$_1$ local environment possesses no symmetry restrictions on the EFG tensor.
We note that in our calculations a misalignment of $\sim$3$^\circ$ between different experiments has been considered.
Besides, we find that the local fields for the U-edge singularities in the SDW phase deviate almost symmetrically from the paramagnetic value by $\sim$\,$\pm$0.05\,T in size and $\sim$\,$\pm2^\circ$ in orientation.
Considering the hyperfine coupling constant $A_{\text{high}}$ this suggests that the amplitude of the modulated magnetic moments is $\sim$0.3\,$\mu_B$, i.e., in agreement with the neutron diffraction results \cite{pregelj2016exchange}.
We stress that the obtained parameters are not exact and should be refined by more involved angular-dependent measurements, yet, they still provide a reasonable estimate of the EFG strength at the O$_1$ site.
Finally, we point out that in contrast to the results at 9.4\,T, the U-shape of the spectral lines at 2.35\,T is much less pronounced, as edge singularities are far more dominant, indicating that middle parts of the spectrum associated with the reduced static magnetic moments are suppressed.
This can be explained by enhanced $T_2$ relaxation, driven phason excitations of the SDW order \cite{zumer1981nuclear, blinc1981magnetic}, that have for small wave vectors (for the NMR case) most pronounced effect on the reduced magnetic moments, i.e., on the middle of the U spectrum.

At the transition into the spin-stripe phase, i.e., at $T_{N2}$, the spectrum dramatically changes.
The signal is almost completely lost, reflecting even stronger $T_2$ relaxation, which corroborates the dynamical nature of the spin-stripe phase on the MHz time scale \cite{pregelj2019elementary}.
We note that the spectra in Fig.\,\ref{fig-100}(a) were measured with the shortest possible $\tau$\,=\,20\,$\mu$s allowed by our spectrometer.
On further cooling, some spectral features reappear, exhibiting an extended array of singularities at the lowest accessible temperature of 1.7\,K.
However, this temperature point lies at the boarder between the spin-stripe and the VC phase, so the resulting spectrum does not fully represent the magnetic order in the VC phase.
Hence, based on the obtained spectrum we cannot learn much about the VC phase, except for the fact that the $T_2$ relaxation in this phase is slowing down, supporting the static long-range order in this phase.

\begin{figure}[!]
\centering
\includegraphics[width=\columnwidth]{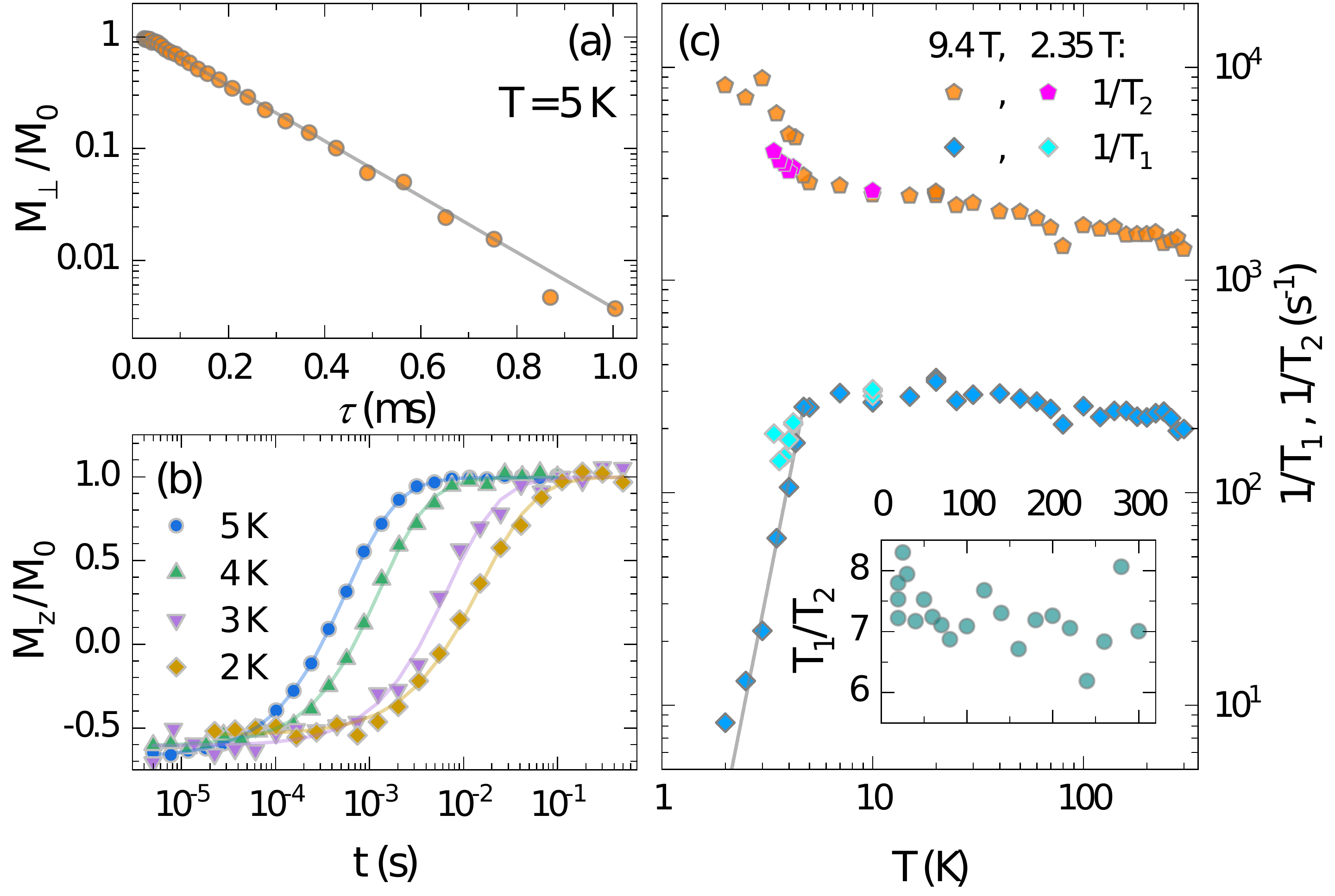}
\caption{A representative low-temperature (a) $T_2$ and (b) $T_1$ measurements (symbols), i.e., transverse- and longitudinal-magnetization ($M_\perp$ and $M_z$) decays, respectively, for the highest-frequency satellite of the only-remaining low-temperature O site measured at 9.4\,T and the corresponding fits (lines; see text). (c) The temperature dependence of the spin-lattice and the spin-spin relaxation rates, $1/T_1$ and $1/T_2$, respectively, measured for the highest-frequency satellite. The solid line shows a fit of the $1/T_1$\,$\propto$\,$T^5$ model. Inset: the temperature dependence of $T_1/T_2$.}
\label{fig-T1T2}
\end{figure}
Next, we explore the spin dynamics through the temperature dependencies of the spin-lattice and the spin-spin relaxation rates, $1/T_1$ and $1/T_2$, respectively.
Both relaxation rates were measured for the outer satellite line [marked by the blue arrow in Fig.\,\ref{fig-400}(a)], i.e., the line corresponding to the $m$\,=\,$-$5/2 to $m$\,=\,$-$3/2 transition [Fig.\,\ref{fig-T1T2}(c)].
Above the magnetic transition, the nuclear magnetization curves [Fig.\,\ref{fig-T1T2}(a) and (b)] yield well-defined and almost temperature independent $1/T_1$ and $1/T_2$, as expected for a paramagnet with strong antiferromagnetic correlations \cite{moriya1956nuclear}.
In fact, the temperature independence of the $T_1/T_2$\,$\approx$\,7 [inset in Fig.\,\ref{fig-T1T2}(c)] ratio suggests that the $T_2$ relaxation is driven by fluctuations of nearby electronic spins, i.e., by the Redfield process \cite{horvatic2002nmr,tokunaga2015reentrant}, yielding
\begin{equation}
\begin{split}
T^{-1}_{2z} = &\frac{1}{2} (T^{-1}_{1x} +T^{-1}_{1y} - T^{-1}_{1z})|_{\omega=0} \\
&+[I(I+1)-m(m-1)-1/2]T^{-1}_{1z}(\omega).
\label{Redfield}
\end{split}
\end{equation}
Taking $m$\,=\,5/2, as the relaxation rates were measured on the outer satellite, the second term in Eq.\,(\ref{Redfield}) yields a prefactor 4.5, which suggests that the $\omega=0$ contribution [the first term in Eq.\,(\ref{Redfield})] is rather anisotropic, implying $T^{-1}_{1x} +T^{-1}_{1y}$\,$\approx$\,$6T^{-1}_{1z}$.
This can be due to the anisotropy of the spin fluctuations as well as that of the hyperfine couplings, since both contributions are allowed by a low symmetry of the O$_1$ site.
In addition, the measurements at 2.35\,T [Fig.\,\ref{fig-T1T2}(c)] show that relaxation times at 10\,K are the same as those obtained at 9.4\,T.
This indicates that in the paramagnetic state relaxation processes are nearly frequency independent, which is indeed expected in the extreme-narrowing limit ensured by strong antiferromagnetic correlations.

At $T_{N1}$, the temperature-independent trend is abruptly broken, as on further cooling $1/T_1$ decreases proportional to $T^p$ with $p$\,=\,5.0(2) for 9.4\,T [solid line in Fig.\,\ref{fig-T1T2}(c)].
The $T^5$ dependence of $1/T_1$ has been observed in several quasi one-dimensional spin systems \cite{mayaffre2000nmr, jeong2017magnetic} and is, in fact, predicted for a three-magnon process for $T\gg\Delta$, where $\Delta$ is the excitation gap of magnons \cite{beeman1968nuclear}.
The fact that neutron scattering experiment in the absence of the magnetic field revealed a magnon gap of $\sim$7\,K (0.6\,meV) (see supplementary information to Ref.\,\onlinecite{pregelj2019elementary}) suggests that this gap is almost completely closed by the applied magnetic field of 9.4\,T ($\sim$6.3 K), pushing the system into the limit $T\gg\Delta$ in agreement with our NMR experiment.
On the other hand, $1/T_2$ below $T_{N1}$ gradually increases, which indicates slowing down of the spin-spin relaxation due to the establishment of long-range magnetic ordering.
We note that the data collected at 2.35\,T exhibit less pronounced temperature dependencies than at 9.4\,T and are most likely affected by the proximity of the $T_{N2}$ magnetic transition [Fig.\,\ref{fig-PD}(a)], occurring less then 1.5\,K below $T_{N1}$.
Moreover, at 2.35\,T we were unable to consistently measure relaxation times below $T_{N2}$, as there is no distinct spectral feature [Fig.\,\ref{fig-100}(a)] that would allow for reliable measurements at the same NMR transition.
At last, we point out the absence of critical fluctuations in the paramagnetic state just above the $T_{N1}$ transition.
Considering a low symmetry of the O$_1$ site [inset in Fig.\,\ref{fig-PD}(a)]\cite{meunier1973oxyde}, a symmetry-based filtration of critical fluctuations is highly unlikely thus implying that the critical region is very narrow, i.e., in line with a three-dimensional nature of the $T_{N1}$ magnetic transition.

\begin{figure}[!]
\centering
\includegraphics[width=\columnwidth]{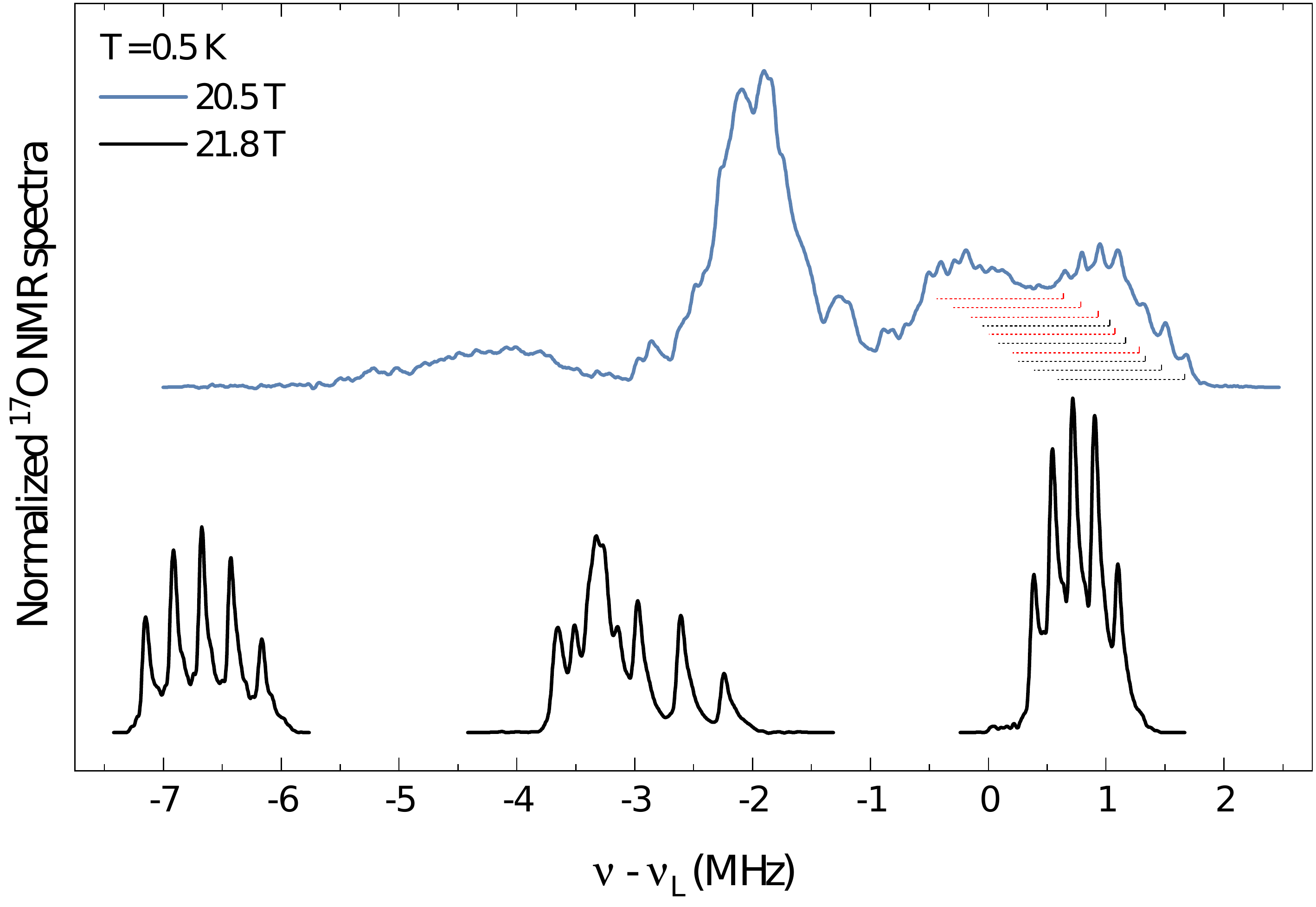}
\caption{$^{17}$O NMR spectra measured in high-magnetic fields in the incommensurate HF phase and in the saturated state.}
\label{fig-HF}
\end{figure}
Finally, we performed $^{17}$O NMR also in high magnetic fields at 0.5\,K.
We measured NMR spectra in the HF phase, i.e., at 20.5\,T, as well as above the magnetization saturation, at 21.8\,T (Fig.\,\ref{fig-HF}). 
This complements the data from our previous study \cite{pregelj2020thermal}, where spectra were measured at fixed frequency by sweeping the magnetic field.
Clearly, the $T_2$ relaxation at these fields is significantly reduced, as in the saturated phase all four quintets are observed again.
Three quintets are found below the Larmor frequency, $\nu_L$, while one quintet is found above $\nu_L$, in agreement with the response observed in the paramagnetic phase at 9.4\,T.
Moreover, the large separation of the quintets at 21.8\,T allows us to distinguish the difference in the EFG strengths at different oxygen sites, which are responsible for the splittings of the lines within the individual quintets.
In particular, for $B||a$, these splittings amount to $\sim$0.25, 0.13, 0.35 and 0.19\,MHz for the O sites sorted from the lowest to the highest frequency, respectively.
We note, however, that these values reflect the EFG strengths along the applied magnetic field, i.e., along the $a$ axis, and can change significantly for different field orientation.

In the HF phase, however, the contributions from the four O sites overlap.
In particular, the low-frequency part of the spectrum is a complex mixture of several overlapping contributions that cannot be disentangled. 
On the other hand, the highest frequency U-shaped feature [Fig.\,\ref{fig-PD}(b)] is most likely associated with the high-frequency O site  (assigned to O$_1$ site).
In fact, it appears that this part of the spectrum is composed of two sets of overlapping U-split quintets [Fig.\,\ref{fig-PD}(b)], which might develop due to symmetry reduction of the magnetic order in the HF phase.
Namely, the two neighboring chains probably exhibit different alignments of the SDW-modulation eigenaxes, which would lead to different projections of the magnetic moments on the external magnetic field.
Similar ordering was found by neutron diffraction in the SDW phase in the absence of the magnetic field \cite{pregelj2016exchange}, where magnetic moments lie along the $a$ axis on one chain and along the $c$ axis on the other.
The possible similarity between the orders in these two phases is also supported by the fact that both phases exhibit similar magnetic ordering vectors \cite{pregelj2019magnetic}, highlighting the importance of the magnetic anisotropy in this system.

%%%% Summary and conclusions
\section{Summary and conclusions}

By employing $^{17}$O nuclear magnetic resonance (NMR), we have investigated magnetic properties of $\beta$-TeVO$_4$ for $B||a$.
We have determined the hyperfine coupling constants for the two out of four crystalographically different O sites, i.e., the two that exhibit the smallest and the largest frequency shifts, which differ by a factor of $\sim$7.
The shift of one  $^{17}$O quintet is considerably smaller than those of the other three, implying that it probably corresponds to the isolated O$_1$ site, positioned on the top of the VO$_5$ pyramid.
For this site, we have also estimated the local EFG tensor.
In addition, we have determined the critical exponent for the magnetic order parameter, $\beta$\,=\,0.37(3), which is in agreement with the theoretical predictions for 3D spin models, and thus corroborates sizable interchain interactions \cite{pregelj2018coexisting}.
Furthermore, in the spin-stripe phase we have confirmed the presence of strong dynamics in the MHz range that is driven by the unusual low-energy bound-phason excitations emerging due to a weak forth-order exchange coupling term \cite{pregelj2019elementary}.
On the other hand, in the SDW phase, the temperature dependence of the $1/T_1$ relaxation at 9.4\,T exhibits the $T^5$ dependence, implying that the magnon gap in this phase is almost completely closed by the applied magnetic field.
Finally, the U shape of the high-frequency part of the spectrum in the HF phase suggests that the neighboring chains might exhibit different alignments of the magnetic moments, similarly as found in the zero-field VC magnetic ground state. 
Still, detailed understanding of the HF phase calls for future density-functional-theory calculations of the EFG tensors at all O sites, which would allow for more involved modeling of the HF NMR spectrum.
For such calculations the derived hyperfine-coupling constants and EFG parameters should serve as an important point of reference, allowing an independent verification of the calculation precision.

\begin{acknowledgments}

This work has been funded by the Slovenian Research Agency (projects J1-9145, J1-2461, N1-0148 and J2-2513, and program No. P1-0125) and the Swiss National Science Foundation (project SCOPES IZ73Z0\_152734/1).
We acknowledge the support of the LNCMI-CNRS, member of the European Magnetic Field Laboratory (EMFL).

\end{acknowledgments}

\end{document}